\documentclass[12pt]{article}
\usepackage{epsfig}
\begin{document}
\begin{flushright}
\end{flushright}
\vspace{-1.cm}
\begin{flushleft}
\end{flushleft}
\vspace{0.5cm}

\begin{center}
  {\Large {\bf 
  Matter Effects of Thin Layers: 
  Detecting Oil by Oscillations of Solar Neutrinos
  }}\\[1.5cm]
    {Ara N. Ioannisian$^{1,3,4}$ and Alexei Yu. Smirnov$^{1,2}$}\\[0.35cm]

{\em 1) International Center for Theoretical Physics, 34100 Trieste, Italy}\\
{\em 2) Institute for Nuclear Research, RAS, Moscow, Russia}\\
{\em 3) Institute for Theoretical Physics and Modeling, Halabian 34, Yerevan-36, 
Armenia}\\
{\em 4) Yerevan Physics Institute, Alikhanian Br.\ 2, Yerevan-36, Armenia}

\end{center} 
\vskip1.cm \centerline {\bf Abstract} 
\vskip 0.5 cm 

  We consider a  possibility to use the solar neutrinos for studies of
small
  scale structures of the Earth and for geological research. Effects of
  thin layers of matter with density contrast on oscillations of 
  Beryllium neutrinos inside 
  the Earth are studied. We find that change of the  
$^7\!Be$ neutrino flux can reach
  0.1 $\%$  for layers with density of oil and    
  size 20  km. 
  Problems of detection are discussed. 
  Hypothetical method 
  would consist of measuring  the $^7\!Be -$ flux by,  
  {\it e.g.}, large deep underwater
  detector$-$submarine which could change its location. 
\\

  \newpage
  \section{Introduction}

  In geophysics the matter density profile of the Earth   
  is determined  by 
  measurements of  the seismic wave velocity profile. Seismic method is used 
  widely  also for the oil search, in particular, in the decision process
  before a trial well is drilled. This method has, however, shortcomings and
  alternatives are highly appreciated.

  Enormous penetration ability of neutrinos is a source 
  of strong temptation to use them 
in tomography and geological studies
\cite{DeRujula:1983ya},\cite{Ermilova:1988pw},\cite{Nicolaidis:1988fe},
\cite{Nicolaidis:1990jm}.
  The first proposal was based on inelastic
  scattering of high energy neutrinos produced by  accelerators
  \cite{DeRujula:1983ya}. The authors have suggested several methods 
  which use difference of the  neutrino interactions 
  in matter with different chemical composition and density.

  Qualitatively  different proposal is based on the elastic forward
scattering (refraction) of neutrinos  
\cite{Ermilova:1988pw},\cite{Nicolaidis:1988fe},\cite{Nicolaidis:1990jm}. 
Refraction modifies properties of  neutrino oscillations
changing both the oscillation length and  depth  
\cite{Wolfenstein:1977ue}. The
effect depends on density of medium. 
Consequently, an  appearance of layers with
different densities on the way of neutrinos changes  
the oscillation pattern.  In this connection,  
some possibilities to use the  neutrino superbeams as well as 
the beams from neutrino factories have been considered \cite{ohl1}. 

  There is a number of studies of the Earth matter effect on oscillations of
  solar neutrinos. It was marked that the solar 
  neutrinos can give information about large scale density
  distribution inside the Earth. 
  In particular, properties of the mantle and the core
  of the Earth can be studied~\cite{Gonzalez-Garcia:2001dj}.

  The Earth matter effect has been considered 
  for the $^7\!Be$ neutrinos \cite{Bemat} in 
  connection to BOREXINO \cite{BOREXINO} 
  and KamLAND \cite{KamLAND} experiments. It 
  was found that the  effect is
  negligible for the LMA solution, at least for these detectors.\\ 

  Main conditions to use neutrinos for the  
  geological studies and searches for  oil
  and minerals are:

  1)  sensitivity to small scale structures: $ d \le$ 100 km;
    
  2)  possibility to move both  a source  
  and detector of neutrinos, 
  so that  substantial part of outer layers of the Earth mantle can be
  scanned. 

  No realistic proposals which satisfy these 
  conditions have been published so far. 
  In this paper we will consider 
  a possibility which is based on:

  \begin{itemize}

  \item   
  Detection  of the solar $^7\!Be - $ neutrinos:  
  For the  LMA solution of the solar neutrino problem 
  the oscillation length of neutrinos
  with $E_{Be} \approx$ 0.86 MeV is in the  required 
  range l$_\nu$/2 $\sim$ (10 $\div$ 20) km.

  \item
  Measuring the time dependence  of the neutrino flux by  
  deep underwater detector - submarine.

  \end{itemize}

  The paper is organized as follows. In section 2 we  consider 
  the effect of thin layers of matter on neutrino oscillations  
(see \cite{thin} for some related work).   We study dependence 
  of the effect on neutrino parameters 
  and  properties of layers.  In section 3 we consider a possibility 
to identify the effect of cavities.  
  In section 4 we discuss problems of detections. 
  Concluding remarks are given in section 5.

  \section{Effects of thin layers on neutrino oscillations}

  Let us consider a system of two mixed neutrinos with values 
  of oscillation parameters   $\Delta m^2$ and $\sin^2  2\theta$   
  from the LMA solution region \cite{Krastev:2001tv}. 
  (We will comment on the effect of the third neutrino at the end of
this section). 
  In the case of the LMA solution, the electron neutrino produced 
  in the center of the
  sun is adiabatically converted to a combination 
  of the mass eigenstates $\nu_1$ and $\nu_2$  
  which is determined by the mixing angle, $\theta_m^0$, in the production
  point:
          \begin{equation}
          \label{}
\nu_e \ \rightarrow \ \cos\theta_m^0~\nu_1 \ + \sin\theta_m^0 ~\nu_2\ .  
          \end{equation}
  The  angle $\theta_m^0$ is given  by:
          \begin{equation}
          \label{}
          \tan^2 2\theta_m^0 \ = \ 
  \tan^2 2\theta \left( 1-\frac{2V_e^0E}{\Delta m^2 \cos 2 \theta} \right)^{-2}  \ , 
          \end{equation} 
  where $E$ is the  neutrino energy,   
  $V_e^0$ is the matter potential for  neutrinos in the 
  production point.  $V_e = \sqrt{2} G_F \rho N_A Y_e$,  
  $G_F$ is the Fermi coupling constant,  
  $\rho$ is the density of medium, $N_A$ is the Avogadro number   and 
  $Y_e$ is the number of electrons per nucleon; $V_e^0 = V_e(\rho^0, Y_e^0)$.

  Conversion effect should be averaged over 
  the neutrino  production region,   
  and in what follows we  will describe  this averaging  by the  
  effective mixing angle $\bar{\theta}^0_m$. 

  Due to loss of coherence \cite{Dighe:1999id}, neutrinos arrive at
  the surface of the
  Earth as incoherent fluxes of $\nu_1$ and $\nu_2$ 
  with relative admixtures given by 
  $\cos^2 \bar{\theta}^0_m$ and $\sin^2 \bar{\theta}^0_m$ correspondingly.

  Let us consider oscillations inside  the Earth.  
  The probability to find $\nu_e$ in the detector can be written as:
          \begin{equation}
          \label{probnue}
          P \ = \ \cos^2 \bar{\theta}^0_m P_{1e} 
  + \sin^2 \bar{\theta}^0_m  P_{2e} \ = \
  \cos 2 \bar{\theta}^0_m P_{1e} + \sin^2 \bar{\theta}^0_m,
          \end{equation}
  where $P_{1e}$ and $P_{2e}$ are the probabilities 
  of $\nu_1 \to \nu_e$ and  $\nu_2 \to \nu_e$
  transitions in  matter of the Earth correspondingly. 

  Suppose neutrinos cross the cavity with length $d$ and density
  $\rho_d$ which differs from the average density of surrounding 
  matter, $\rho$.  
  That is, the neutrinos cross consecutively three layers: 
  with lengths and densities
  ($l_{1}$, $\rho$), ($d$, $\rho_d$), ($l_2$, $\rho$). 
  The total length of the trajectory is $L = l_1 + d + l_2$.

  Let us find an  effect of the cavity on  oscillations. 
  Introducing, $P^0_{1e}$,  the $\nu_1 \to \nu_e$  oscillation probability  in
  absence of cavity (it corresponds to oscillations in the 
  unique layer  of density $\rho$ and length $L$) we can
  write $P$ as:
\begin{equation}
\label{main1}
P  =   P^0  + \Delta P, 
\end{equation}
\begin{equation}
\label{main2} 
P^0  =  \sin^2 \bar{\theta}^0_m \ + \ \cos 2 \bar{\theta}^0_m P^0_{1e}, 
\end{equation}
\begin{equation}
\label{main3} 
\Delta P \  =  \cos 2 \bar{\theta}^0_m~ (P_{1e} - P^0_{1e}).
\end{equation}
Here $P^0$ is the probability to find $\nu_e$ in the absence of cavity 
and $\Delta P$ is the change of the probability due to the cavity effect.

  We will calculate the effect for the beryllium 
  neutrino line with energy  $E_{Be} = 0.86$ MeV. 
  For such an energy  the matter effect on mixing and 
  oscillations is very weak. It is determined by a small parameter:
          \begin{equation}
          \epsilon \ \equiv \ \frac{2 V_e E}{\Delta m^2 } \ \approx \ 
                2.6\cdot10^{-3} \left( \frac{\rho}{2.7 {\rm g/cm^3}}           \right) 
                        \left( \frac{7\cdot10^{-5} {\rm eV}^2}{\Delta m^2}  \right)
                        \left( \frac{Y_e}{0.5} \right) \   
          \end{equation} 
  which characterizes  deviations of the mixing angle 
and the oscillation length in
  medium from their vacuum values. We will use $\epsilon$ as an 
  expansion parameter and find effects in the lowest order in $\epsilon$. 
  The oscillation length is  determined by $\Delta m^2$: 
          \begin{equation}
          \label{}
          l_\nu^m \approx l_\nu \approx 30.7 \ {\rm km} \  
          \left( \frac{7 \cdot 10^{-5}{\rm eV}^2}{\Delta m^2} \right)  \ .
          \end{equation}

 The probability of $\nu_{1} \rightarrow \nu_e$ oscillations equals 
\begin{equation}
\label{p1e}
P_{1e} \ = \ | \cos \theta  \ S_{11} \ + \ \sin \theta \ S_{12} |^2 \ ,
\end{equation}
  where $S = ||S_{ij}||$ is the evolution matrix in the mass eigenstate
  basis, $\vec{\nu}$. The matrix $S$ can be
  written  as:
\begin{equation}
\label{smatr}
S \ =  \ U D_2 U^{\dagger} U_d D_d U_d^{\dagger} U D_1  U^{\dagger} \ . 
\end{equation}
  Here $U \equiv U(\rho)$ and $U_d \equiv U(\rho_d)$ are  the mixing
matrices of mass  eigenstates in the  layers 1, 2  and
in the  cavity correspondingly. They are defined in such a way that  
$\vec{\nu} = U\vec{\nu}_m$, where $\vec{\nu}_m$ is the basis of 
eigenstates of the Hamiltonean in matter. In (\ref{smatr}),  the diagonal
matrices, 
  \begin{equation}
  \label{dmatr}
  D_i \equiv  diag(1, e^{i \Phi_i}), \ \ \  i = 1, 2, d, 
  \end{equation}
  describe   evolution of the neutrino eigenstates,  
$\vec{\nu}_m$, in layer $i$. $\Phi_i$ is
  the phase of oscillations:
\begin{equation}
\label{phasei}
\Phi_i \ = \ \frac{\Delta m^2 l_i}{2 E} \sqrt{(\cos 2\theta
  -\epsilon_i)^2+\sin^2 2\theta } \ ,
\end{equation}
$\epsilon_i  \equiv \epsilon(\rho_i)$. In absence of cavity we have
\begin{equation}
\label{szero}
        S = S_0 \ = \ U D U^+ \ ,
\end{equation}
where $D = diag(1, e^{i \Phi})$ and 
\begin{equation}
\label{}
          \Phi \ = \ \frac{\Delta m^2 L}{2 E} \sqrt{(\cos 2\theta
  -\epsilon)^2 + \sin^2 2\theta } 
\end{equation}
is the total oscillation phase in absence of the cavity.

Using Eqs. (\ref{p1e},\ref{smatr},\ref{dmatr}), the  
straightforward calculation leads to the $\nu_1 \rightarrow \nu_e$ 
probability in the lowest order in  $\epsilon$:
          \begin{equation}
          \label{owc}
          P_{1e} \ \approx  \ P^0_{1e} \ + \  \epsilon \ \xi \ \sin^2
2\theta \ 
          \sin \left(\Phi_2 + \frac{\Phi_d}{2} \right) \ \sin
  \frac{\Phi_d}{2}~,
          \end{equation}
where 
\begin{equation}
\xi \equiv  \frac{(Y_e  \rho) - (Y_{ed} \rho_d)}{Y_e \rho} 
\label{contrast}
\end{equation}
is the density contrast. 
In absence of cavity  the probability equals 
\begin{equation}
\label{owoc}
P^0_{1e} \ = \ \cos^2 \theta - \epsilon \sin^2 2\theta \ 
\sin^2 \frac{\Phi}{2} \ .  
\end{equation}
  %
And  according to eq. (\ref{main3}) and (\ref{owc})
the effect  of cavity can be written as 
  \begin{equation}
  \label{fin}
  \Delta P   =   \epsilon \xi \cos 2\bar{\theta}^0_m 
  \sin^2 2\theta \sin\left(\Phi_2 + \frac{\Phi_d}{2}\right) 
  \sin \frac{\Phi_d}{2}~.
  \end{equation}

  The following remarks are in order:

  According to eq. (\ref{fin}) the effect of cavity is proportional 
  to expansion parameter $\epsilon$,  the density 
  contrast, $\xi$, 
  and $\cos 2 \bar{\theta}^0_m$. For maximal mixing in the
  production region the effect would be  zero. For the best fit 
  point of the LMA solution  
  we get $\cos 2 \bar{\theta}^0_m \approx 0.3$.

Important feature of the result (\ref{fin}) is that $\Delta P$ 
does not depend  on the oscillation effect in the first layer 
(before cavity), in particular,  it does not depend on the phase  $\Phi_1$.  
In contrast, $\Delta P$ does depend
  on $\Phi_2$ -- the oscillation phase in the layer between the cavity and
  detector. This 
  property appears in the lowest order in $\epsilon$ and 
  is related to the fact that
  initial state is the mass eigenstate and final state (at the detector) is
  the flavor state. It can be shown that 
  if initial state is  $\nu_e$ and final
  state -- $\nu_1$, the probability 
  $\Delta P$ does  not depend on
  $\Phi_2$ (phase between cavity and detector) 
  but will depend on $\Phi_1 -$  the
  phase acquired  in the first layer. 
Thus,  the interchange
  of $\nu_1$ and $\nu_e$ in the initial and final states leads to change
  $\Phi_2 \to \Phi_1$. 
  This feature is important for identification of the cavity effect  
  (see section 3).\\ 

  Let us estimate the size of possible effect 
  depending on parameters of neutrinos and the cavity. 
The cavity produces the  change of the event rate during the time 
when the neutrino beam crosses the cavity.  
The relative value of the change  $\Delta N / N = \Delta P /P$ equals in the lowest order in 
$\epsilon$ 
          \begin{equation}
          \label{releff}
          \frac{\Delta P}{P^0} \ = \ f(\Delta m^2, \theta) \ \epsilon \ \xi
          \sin\left(\Phi_2+\frac{\Phi_d}{2} \right) \sin \frac{\Phi_d}{2}\ , 
          \end{equation}
  where
          \begin{equation}
          \label{func}
          f(\Delta m^2, \theta)  \ = \ 
          \frac{2 \ \cos 2\bar{\theta^0_m} \ \sin^2 2\theta}
          {1 \ + \ \cos 2 \bar{\theta^0_m} \cos 2 \theta }.  
          \end{equation}
Here we have taken $P^0 = (1 + \cos 2 \bar{\theta}^0_m \cos 2\theta)/2$ 
neglecting the Earth matter effect. 
The dependence of $f(\Delta m^2, \theta)$ on $\tan^2 \theta$ 
for $\Delta m^2$ from the allowed LMA region is shown in Fig.~1.
According to Fig.~1 the function $f(\Delta m^2, \theta)$ 
has maximum  $f^{max} \sim 0.4 $ at about  the best fit value of 
$\tan^2 \theta$.

\noindent
\vskip 1cm
\centerline{\epsfxsize=90mm\epsfbox{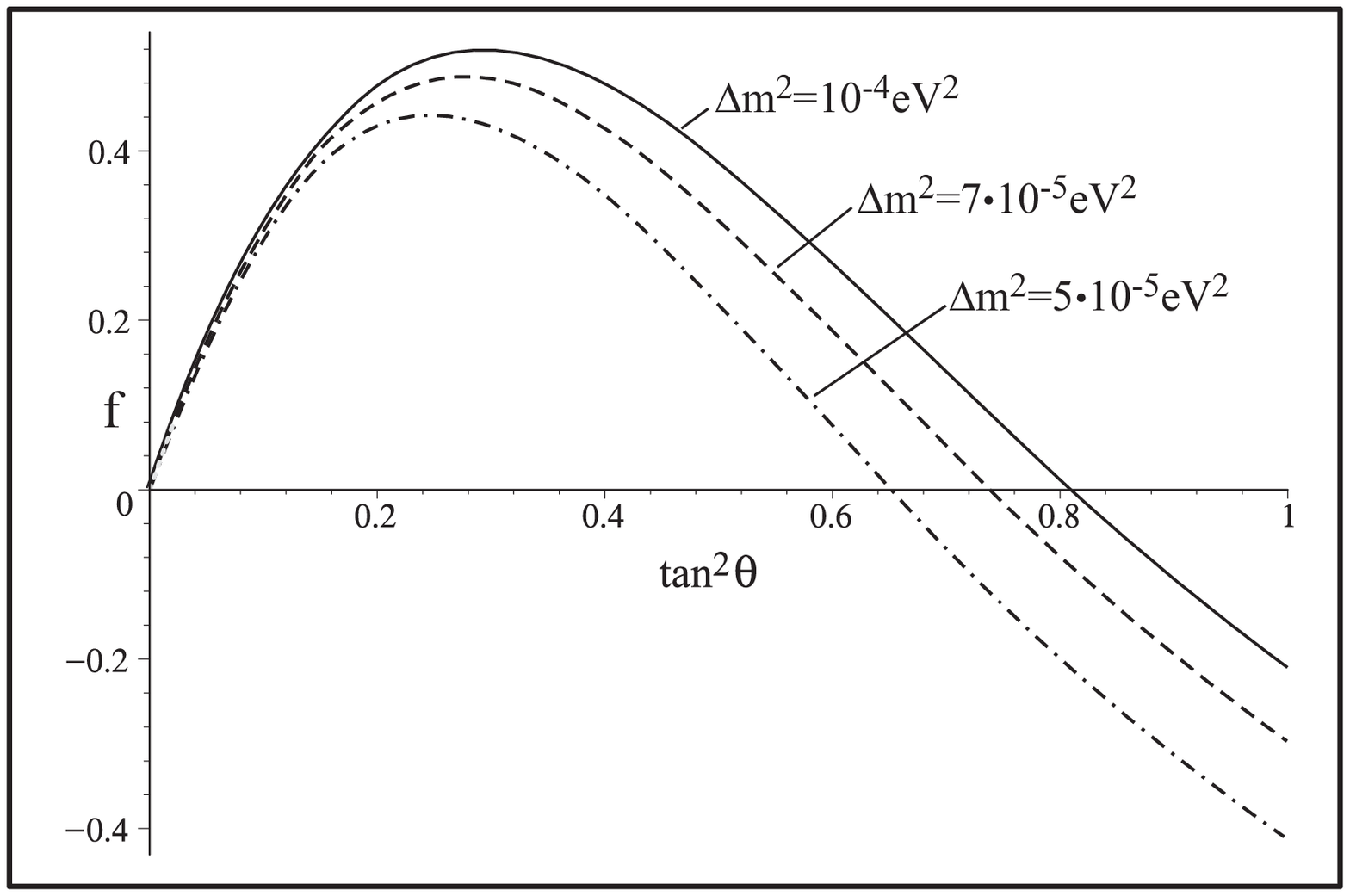}}
\noindent
Fig.\ 1. \ The dependence of $f(\Delta m^2,\theta)$ on
$\tan^2 \theta$ for  $\Delta m^2 \ = \ 5 \cdot 10^{-5}$ eV$^2$, 
$ 7 \cdot 10^{-5}$ eV$^2$ and
$10^{-4} $eV$^2$. 
\vskip 1cm 

The relative change of the probability can be rewritten as 
\begin{equation}
\label{dPoP}
\frac{\Delta P}{P^0} \ =  A_{max} \sin
  \left(\Phi_2+\frac{\Phi_d}{2}\right) \sin
          \frac{\Phi_d}{2}~,   
  \end{equation}
where 
\begin{equation}
\label{amax}
A_{max} \ \equiv \ \left.\frac{\Delta P_e}{P_e}\right|_{max} \ = 
\ \epsilon \ {\xi} \, f(\Delta m^2,\theta)  
\end{equation}
is  the maximal  value of $\Delta P/ P^0$ which corresponds to both sines 
being 1.
The effect increases with decrease of $\Delta m^2$.
For minimal allowed value $\Delta m^2 = 5 \cdot 10^{-5}$eV$^2$, and  
$\tan^2 \theta =0.3$,  $\xi=0.64$  we get
\begin{equation}
\label{amaxnum}
A_{max} = \ 0.1\% . 
\end{equation}
In Fig.~2 we show $A_{max}$ as the function of
$\Delta m^2$ and $\tan^2 \theta$.
The effect weakly depends on  $\theta$ in the allowed LMA region. \\

\noindent
\vskip 1cm
\centerline{\epsfxsize=110mm\epsfbox{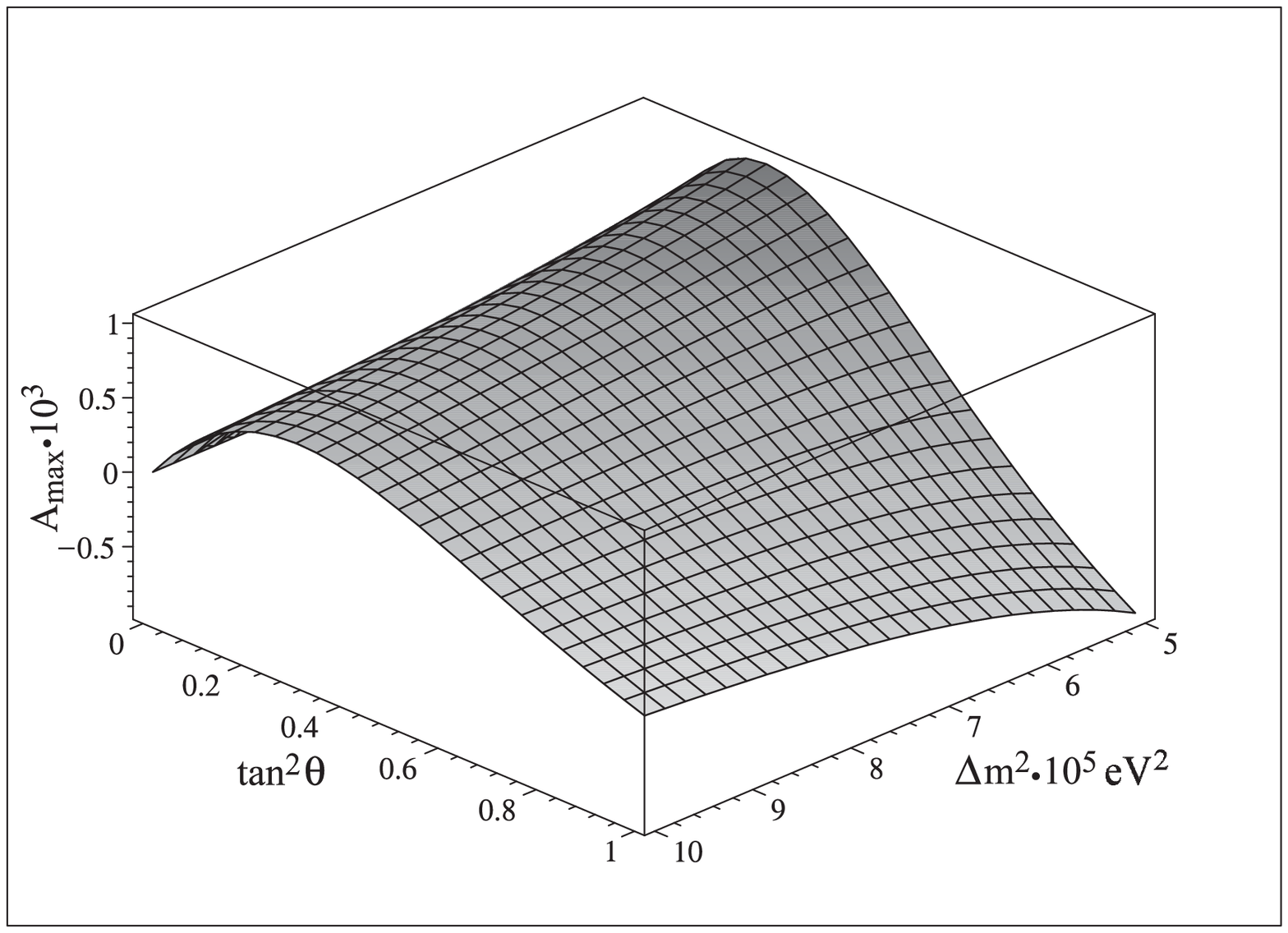}}
\noindent
Fig.\ 2. \ Maximum value of the relative effect, 
$A_{max} \equiv (\Delta P / P)_{max}$, 
as the function of $\Delta m^2$ and $\tan^2 \theta$.
  \noindent
\vspace{1cm}

  If detector is placed underwater, neutrinos 
  will cross an additional layer of water,
  $l_W$,which can be rather large.
  Straightforward calculations lead to result, 
  similar to eq.(\ref{releff}) in which
  $$
  \Phi_2 \to \Phi_2' =  \Phi_2 + \Phi_W, ~~~\Phi_w \simeq \frac{\Delta
  m^2l_W}{2E}, 
  $$ 
  where  $\Phi_w$ is the phase acquired by neutrinos in
  water. \\

Let us comment on possible effect of the third neutrino. We consider the 
scheme in which $\nu_3$ is separated from two others by the mass 
gap $\Delta m^2_{23} = \Delta m^2_{atm} =  (2 - 3) \cdot 10^{-3}$ eV$^2$. 
(So, the scheme explains the atmospheric neutrino data.)
We assume that this neutrino has small admixture of the electron neutrino 
described by $U_{e3} < 0.15$. The influence of matter on this mixing 
is determined by $\epsilon_{atm} = \epsilon (\Delta m^2_{atm}) \sim 7
\cdot 10^{-5}$. 
Consequently, variations of the flux due to cavity effect should be  
smaller than $7 \cdot 10^{-5}$. 
In fact, the observable effects are further suppressed by smallness of 
$U_{e3}$ and averaging of oscillations associated with the 
third neutrino. For $\Delta m^2_{atm}$ the oscillation length is smaller than  
$1$ km. Also interference between the modes of oscillations driven by 
$\Delta m^2_{atm}$ and solar $\Delta m^2$ produces negligible
effect.  So, we conclude that the cavity effect is mainly due 
to oscillations driven by solar $\Delta m^2$.

  \section{Effect of cavity. Exposure time}

Let us consider the time dependence of the electron neutrino flux in,  
{\it e.g.},  deep underwater detector.   
For fixed position of the detector, the time dependence 
is due to rotation of the Earth.  
The rotation leads to  certain change of the trajectory 
of solar neutrinos inside the Earth,  
and consequently, to modification of  the oscillation effect. 
In general,  the change  of signal with time is both due to the cavity
effect and  the effect of regeneration in $P^0$ (see (\ref{main2})). 
The problem is to disentangle  these two effects. 
 
Let $t_{exp}$ be the exposure time - the time during 
which the trajectory of solar neutrinos crosses the cavity. 
The effect of cavity consists of an additional 
change of the neutrino flux during 
exposure time $t_{exp}$ on the top of usual regeneration effect.  

According to (\ref{main2}) and  (\ref{owoc}) the probability in the absence of
cavity equals  
  \begin{equation}
  \label{regen}
  P^0(t)  =  \sin^2 \bar{\theta}^0_m   + 
  \cos 2\bar{\theta}^0_m \left[\cos^2 \theta - \epsilon(t) \sin^2 2\theta  
          \sin^2 \frac{\Phi(t)}{2} \right].  \\
  \end{equation} 
  In principle, this probability can be predicted with high enough 
  accuracy. Indeed,  the average radial density profile 
\cite{Dziewonski} is a subject to local
variations at the level $(5- 7)\%$ \cite{Geller:2001ix}.  
Therefore one expects  $\sim (5 - 7)\%$ variations of 
$\epsilon(t)$. This uncertainty can  be important for very small 
  values of the density contrast: 
$\xi \sim (5 - 7)\%$ or very small phases in Eq. (\ref{releff}). 
Thus, one should look for the deviation 
of the time dependence of  signal from  the signal  
expected according to (\ref{regen}).  
  Identification of the cavity effect  depends on specific 
  conditions of observation. If the cavity is close enough to the 
  detector, the distance $l_1$, and therefore 
  $L$,  can change significantly during the exposure time,  
  so that  the averaging over the phase $\Phi$ will occur. 
  As a result 
\begin{equation}
\label{owoc1}
          \overline{P^0} \ = \ 
\sin^2 \bar{\theta}^0_m +  \cos 2\bar{\theta}^0_m
\left(\cos^2 \theta - \frac{\epsilon}{2} \sin^2 2\theta\right) \ 
\end{equation} 
practically does not depend on time, and whole time 
dependence is due to presence of cavity.\\

Due to smallness of relative change of the probability,
$\Delta P / P$, one needs to collect the integral effect during long time of
observation.  Therefore,  
the effect of cavity  is determined by the relative change 
of probability averaged over the exposure time:
\begin{equation}
\label{averprob}
\frac{\overline{\Delta P}}{P} \ = \ \frac{1}{t_{exp}} A_{max} \int dt 
~\sin \left(\Phi_2' (t) + \frac{\Phi_d(t)}{2}\right) 
\sin \frac{\Phi_d(t)}{2} \ .
\end{equation} 
In general, the cavity has complicated form, 
so that $d = d(t)$ can change  with time
during exposure interval significantly. 
Averaging over $d$ gives
\begin{equation}
\label{d-aver}
\frac{\overline{\Delta P}}{P} \ = \ \frac{1}{t_{exp}}\frac{A_{max}}{2} 
\int dt \cos \Phi_2'(t) \ .
\end{equation} 

The size of the cavity should not  be  much smaller than the
oscillation length. For the $\Delta m^2 \simeq 7 \cdot 10^{-5}$eV$^2$ 
the oscillation length is  30 km. Therefore maximum effect should be seen
for the cavity size $l_\nu / 2 \simeq
15$ km. For  smaller cavities  the effect will be suppressed according
to the Eq. (\ref{releff}). \\

$\Delta P$ is the periodic function of the phase $\Phi_2'$.
Averaging over $\Phi_2'$  leads to disappearance of the cavity
effect $\overline{\Delta P}=0$. 
Suppose  averaging is absent. 
By moving the detector one can vary 
distances $l_2$ and $l_W$.   
Selecting  the distance $l_2$ and  $l_W$ in such a way that 
$\Phi_2' + \Phi_d/2  = \pi/2 + \pi k$ ($k=$ integer) 
one can maximize the effect
\begin{equation}
          \label{}
 \Delta P  \ = \ \epsilon \ \xi \ \sin^2 2\theta  \cos 2 \bar{\theta}^0_m 
          \sin \frac{\Phi_d}{2} \ .
          \end{equation}
Correspondingly, there are zones  of $l_2$  with  high and low
sensitivity. Moving the detector by $l_{\nu}/2$ one can cover whole 
space.\\

The width of the  beryllium neutrino line, $\Delta E_{Be} \approx 1.5$ 
keV \cite{bahcall}, is very small. Consequently, no averaging 
over the energy occurs. 
Averaging  would require the length of trajectory: 
$l_1 > l_{\nu} E_{Be}/\Delta E_{Be} \sim 2 \cdot 10^{4}$ km which 
is larger than the Earth diameter. \\

  The sign of the effect depends on the sign of density  contrast,  the distance
  between the cavity
  and the detector, $l_2$, and the size of the cavity itself. 
In particular, if the
cavity is close enough to the detector, so that 
$\Phi_2 \ll \pi$,  and therefore $\Delta P \sim \xi \sin^2 \Phi_d/2$, 
the $\nu_e$ flux (signal) increases, when neutrinos cross the 
cavity  for the positive contrast $\xi > 0$ $\rho_d < \rho_2$ (which is satisfied for 
oil or  water), and it decreases for the
negative contrast. Similar situation is for $\Phi_2 \sim 2\pi k$, 
$k = 1, 2, ... $. The sign of the effect changes if the detector approaches 
or removes from the cavity by the distance $\sim l_{\nu}/2$.

  Detecting the cavity from the first  position of the detector one
  determines the line
  along which the cavity is situated. Observing the cavity from two different
  positions of detector one can reconstruct location of the cavity
  unless another cavity is detected from the second position.
  The latter can be checked by observation from the third position of the
  detector.

In general, one should take into account  finite size of the region in
which beryllium neutrinos are produced inside the sun. The diameter of the
production region ($\sim 0.2 R_{\odot}$) has the angular size at the Earth
$\theta_{Be} = 9 \cdot 10^{-4}$ radian. 
Therefore, the cavity with transverse size
$h$ shields the production region completely if 
          \begin{equation}
          \label{}
          l_2 \le {h \over \theta_{Be}}  
          \ \simeq \ 1100 \, {\rm km} \, \left( {h \over 1 {\rm km}}
  \right) \ .
          \end{equation}

If  the cross-section of the cavity, $S_d$, is smaller 
than $(l_2 \theta_{Be})^2$
an additional suppression factor appears 
in the size of the  relative change of the flux:
\begin{equation}
\frac{\Delta P}{P^0} \ \propto \ \frac{S_d}{(l_2 \theta_{Be})^2}. 
\end{equation}
  
The integral effect is determined by the {\it exposure time}, which in turn
depends on  (i) the size and shape  of the cavity, 
(ii) its location, (iii) position of detector and 
(iv) season of the year. 
For simplicity we will consider the cavity of the parallelepiped  
form situated in the horizontal layer,  with  
$h$ being the thickness in the vertical plane, $l_{w} -$ the  width in
the horizontal plane, and as before, $d-$  the size of the cavity in 
horizontal plane along the direction from the Sun  to the detector.   

Let  $t_{exp}^{day}$ be the exposure time during the day,    
that is, during one crossing of the cavity by the neutrino trajectory.
If $\nu-$ trajectory crosses the cavity $N$  days during a
year the total exposure time during the year equals:
\begin{equation}
t_{exp}  \ \sim \ t_{exp}^{day}\cdot N. 
\end{equation}

Let us estimate  the exposure time for different 
possible positions of the cavity. 

Suppose the  cavity is in the  equatorial region. If the detector is situated
in the same horizontal plane as the cavity, 
the diurnal exposure time is determined by 
the cavity thickness:   
          \begin{equation}
          \label{}
          t_{exp}^{day} \ \sim \ {t_{day} h \over 2 \pi l_2 }  
  \ \simeq \ 2.3   
          \, {\rm min}     
          \left( {h \over 1 {\rm km}} \right)  
          \left( {100 {\rm km} \over {l_2}} \right)  \ , 
          \end{equation}
where $t_{day} \equiv 24$ hours. If the detector is out of the plane or
cavity is inclined, 
the exposure time can be larger. However, at the same time  
the length of $\nu$ trajectory in the cavity will 
be shorter. Both factors can compensate each other  
in the  integral number events.

The cavity will shield the detector during $N =  365$  
days, so that   the  total exposure 
time during the year can be as large as $\sim 14$ hours.

  Let us consider the cavity near  the North pole. Here the search will be
  possible in spring or in
  autumn when the sun appears  at horizon. At this time the sun  rises
  by $\sim 7 \cdot 10^{-3}$ radian/day. Therefore the sun will be 
shielded by the cavity during $N = h/(7 \cdot 10^{-3} l_2)$ days.  
For thickness $h =  1$ km we get $N = 1.5$ days. The diurnal exposure 
is determined by the width $l_{w}$, so that total exposure time equals 
\begin{equation}
          \label{texpartic}
          t_{exp} \ = \ {l_{w} t_{day} N \over 2\pi l_2 } day \simeq
          1 \, {\rm hour} \, 
          \left( {l_{width} \over 20 {\rm km}} \right)
          \left( {100 {\rm km} \over l_2} \right)  
           \left( {h \over 1 {\rm km}} \right) \ ,          
\end{equation}

Let us consider now the search 
at latitudes 67$^0$ which correspond to the polar circle. In
this case in December the sun is at horizontal (south) direction  
for 18 days a year. The diurnal exposure time is the same 
as  in previous case. 
As a result, $t_{exp} = 13.5$ hours.  
Similar consideration holds for observation in June 
at the south latitude 67$^0$. \\

\section{Detection}

Clearly, detection of so small ( $< 0.1 \%$) effect 
at low energies and during  
restricted time interval is a very challenging task.
It may require new technological developments. 
Here we will give simple  estimations which  
help to understand problems of the method.  

Let us consider a possibility to detects the effect of the cavity by 
large underwater detector-submarine.  

The background conditions should be as good as in the underground
detectors of Beryllium neutrinos \cite{BOREXINO,KamLAND}. 
To suppress the background 
generated by the cosmic rays,  the detector 
should be placed deep underwater $-$ at about 3500 m below sea level. 
This certainly restricts applications of the method and makes it more 
difficult  technologically. 
Indeed, it is unlikely that, {\it e.g.}, oil reservoirs are very deep 
although   such possibility is not excluded \cite{BE}. On the other hand,
there is  not too many  places where the sea becomes very deep near the cost. 

The background and therefore the depth could be reduced for detectors 
which use  light  elements ({\it e.g.},  helium) and efficient  
anti-coincidence shielding.

Let us estimate the  change of number of events 
due to presence of cavity and  required size of the detector.
  For definiteness we will consider detection of neutrinos
  via the  $\nu e^-$scattering in  scintillator.
  The rate of events due to the Beryllium neutrino flux 
  without oscillations 
  is \cite{BOREXINO}:
\begin{equation}
          r_0 =  20 {\rm{events \over kt \cdot hour}}~. 
\end{equation}
Then in presence of neutrino conversion  the expecting rate is
          \begin{equation}
          r = \left[P + {1 \over 5}(1 - P)\right] r_0~, 
          \end{equation}
  where $P$ is the survival probability of the electron neutrinos. 
  Here  we have taken into  account that
  $\nu_e e^-$ cross section is about 5 times larger than $\nu_\mu e^-$
  cross section for $^7 Be$ neutrino scattering with  energy transfer
  (0.25 -  0.664)  MeV. 
  Taking typical value $P^0  \sim 0.6$ we get
          \begin{equation}
          r =  14 {\rm {events \over kt\cdot  hour}}.
          \end{equation}

Total number of events detected during the year  equals
\begin{equation}
          N =  r \cdot M_d \cdot t_{exp}^{year}, 
\end{equation}
where $M_d$  is the mass of detector.

The effect of cavity consists of the increase (or decrease)
of the number of events by amount
\begin{equation}
\Delta N_c = N \cdot  \overline{\frac{\Delta P}{P}} \cdot 
\frac{1}{1 + 1/4P^0} = 0.7 N \cdot \overline{\frac{\Delta P}{P}}
\label{ncav}
\end{equation}
(see Eq. (\ref{averprob})). Last factor in this equation 
reflects dumping  effect due to contribution of neutral 
current interactions. 
This factor is absent for detection 
via the charged current interactions. 
     
  Let us find the number of events needed to establish an 
  excess (or deficit)  of the number of event due to
  presence of the cavity at   $k \sigma$ level ($k$ is integer).
  We take an ideal situation of 100 \% efficiency of detection,
  absence of background,  absence of averaging,  etc..
  In this case the possibility to identify the effect of cavity will be
  determined  by statistical fluctuation of the neutrino signal 
during the exposure time:
  $\Delta N_{stat} = \sqrt{N}$. The effect of cavity should be larger than
  the fluctuation. The effect of cavity at $k\sigma$ level  will be achieved 
if
  $\Delta N_c \geq  k \Delta N_{stat} = k \sqrt{N}$. From this
  condition we find using Eq. (\ref{ncav}):
          \begin{equation}
          N  \geq  2 \cdot k^2 \cdot \left(\overline{\frac{\Delta
  P}{P}}\right)^{-2}.
          \label{nevent}
          \end{equation}
  For $k = 3$ ($3 \sigma-$ level)  and ${\Delta P}/{P} = 0.001$ 
  this inequality gives 
          \begin{equation}
          N  \geq  1.8 \cdot 10^7 {\rm events}. 
          \label{nevent1}
          \end{equation}

Consequently,  the mass of detector should be
          \begin{equation}
          M_d \geq \frac{N}{r t_{exp}} = \frac{2 \cdot k^2}{r t_{exp}}
          \cdot \left(\overline{\frac{\Delta P}{P}}\right)^{-2}. 
          \end{equation}
  Taking the  exposure time $t_{exp} \sim 10$ hours
  we find $M = 130$ Mt. In general,
          \begin{equation}
          M_d = 130 {\rm Mt} \frac{k^2}{9} \frac{10 {\rm hours}}{t_{exp}} . 
          \label{vol}
          \end{equation}

  For the density of scintillator $0.8$ g/cc 
  a detector could have dimensions
  $\sim$(500 m)$^3$. 
(Compare with SuperKamiokande which 
has diameter 39 m and the height 42 m.)
  Notice that for such a large detector the problem can appear with
  fiducial volume. In our estimations of the exposure time
  we have neglected the finite sizes of the detector.
  In fact, only part of the so large detector can be shielded by
  the cavity: that is, the trajectories of neutrinos which cross the
cavity  will cover only part of the detector during all exposure time. 
Therefore the effect of cavity will be further suppressed.

We have found that the  size of  detector
  (even in the most favorable situation) should be an order
  of magnitude larger than the size of future underground detectors
  which are now under consideration \cite{uno}. Solution 
  of this problem may  require
  next step of  technological  developments.

  \section{Discussion and conclusions}

We have studied effects of the thin layers  of matter 
on oscillations of solar neutrinos, in particular, on the Be-neutrinos. 
The effect is proportional to $\epsilon$, the effective 
mixing  parameter $\cos 2 \bar{\theta}_{m}^0$ in the production point, 
and the density contrast.  

The oscillations occur in the vacuum dominating regime and 
the matter effect appears as the small correction 
to the vacuum  oscillations pattern.  
For the beryllium line the oscillation length is about  20 - 40 km 
and the matter effect parameter $\epsilon \sim 3 \cdot 10^{-3}$. 
This parameter put the absolute upper bound on the effect for a given energy. 
The only possibility to further enhance the effect is to 
have  several layers with  configurations which  
add up their effects  coherently \cite{param}.

The interesting feature of the oscillation result is that for solar neutrinos 
it does not depend on the oscillation  phase acquired by the neutrinos before the layer with 
density contrast. 

We have considered a possibility to study small scale structures 
inside the Earth via oscillations of the  solar Beryllium neutrinos. 
Such a study can be sensitive 
to cavities with density contrast $\sim O(1)$  and  the size (10 - 20) km 
situated not too far ($<$ hundreds km) from the detector.  
The cavity can produce 
a change  of the flux by  about 0.1 \% at most. We have found 
that exposure time during the year can reach about  10 hours. 
So, the detection of a cavity at  $3 \sigma$ level 
would require a detector of about 130 Mt size.

There are some ways to reduce the mass of  detector: 

  1) The exposure  time can be increased 
  if the detector moves following the neutrino
  trajectories which cross the cavity. For this, of course, some
  {\it a priori} knowledge of possible position of the cavity is needed.
  
  Another possibility is to perform observations in the same place during 
several years.

  2) If sterile neutrinos participate in  oscillations one can 
  use  coherent effects on scattering of neutrinos 
  on nuclei by neutral currents in order
  to increase the cross section \cite{stodolsky}.

Refraction effect increases with neutrino energy: 
$\epsilon \propto E$. For the boron neutrinos we get 
$\epsilon \sim 0.03$.  
Moreover, the cross section 
of the $\nu e - $scattering  is proportional to neutrino  energy. 
Finally, the background conditions are better. 
So, one may think to use 
boron neutrinos instead of beryllium neutrinos. 
However, the flux of boron neutrinos is 
3 orders of magnitude smaller,  and  
therefore total number of events  turns out to be 
one order of magnitude smaller than  in the case of beryllium neutrinos. 
There are other problems: the oscillation length is now 
about 300 km, and therefore  with boron neutrinos 
 one can probe larger structures  than with $Be-$ neutrinos  
Furthermore, due to continuous spectrum  the problem 
appears with averaging over energies of neutrinos.

Further studies are needed to show  feasibility of this method 
to study of small scale structures inside the Earth and for the geological research.

  

\begin{thebibliography}{99}

  \bibitem{DeRujula:1983ya}
  A.~De Rujula, S.~L.~Glashow, R.~R.~Wilson and G.~Charpak,
  Phys.\ Rept.\  {\bf 99}, 341 (1983).

  \bibitem{Ermilova:1988pw}
  V.~K.~Ermilova, V.~A.~Tsarev and V.~A.~Chechin,
  {\it  SOV. PHYS. LEBEDEV INST. REP. (1988) NO. 3 51-54. (KRATKIE
  SOOB. FIZ. (1988) NO. 3 38-40)}.

  \bibitem{Nicolaidis:1988fe}
  A.~Nicolaidis,
  Phys.\ Lett.\ B {\bf 200}, 553 (1988).

  \bibitem{Nicolaidis:1990jm}
  A.~Nicolaidis, M.~Jannane and A.~Tarantola,
  J. Geophys. Res. {\bf 96}, 21811 (1991).


 \bibitem{Wolfenstein:1977ue}
  L.~Wolfenstein,
  Phys.\ Rev.\ D {\bf 17}, 2369 (1978).
  S.~P.~Mikheev and A.~Y.~Smirnov,
  Sov.\ J.\ Nucl.\ Phys.\  {\bf 42}, 913 (1985)
  [Yad.\ Fiz.\  {\bf 42}, 1441 (1985)].
  S.~P.~Mikheev and A.~Y.~Smirnov,  
  Nuovo Cim.\ C {\bf 9}, 17 (1986).

\bibitem{ohl1} T. Ohlsson, W. Winter 
Phys. Lett. {\bf B512} 357, (2001),  
e-Print Archive: hep-ph/0105293;  
T. Ohlsson, W. Winter,  
e-Print Archive: hep-ph/0111247.  

\bibitem{Gonzalez-Garcia:2001dj}
see {\it e.g.}, E. Lisi, D. Montanino, Phys. Rev. {\bf D56} 1792 (1997);   
J.N. Bahcall, P.I. Krastev, Phys. Rev. {\bf C56} 2839 (1997); 
G.L. Fogli, E. Lisi, D. Montanino
A. Palazzo,  Phys. Rev. {\bf D62}  113003 (2000);  
M.~Maris and S.~T.~Petcov,   arXiv:hep-ph/0004151; 
M.~C.~Gonzalez-Garcia, C.~Pena-Garay and A.~Y.~Smirnov, 
Phys.\ Rev.\ D {\bf 63}, 113004 (2001). 

\bibitem{Bemat}
  A.~de Gouvea, A.~Friedland and H.~Murayama,
  JHEP {\bf 0103}, 009 (2001)
  [arXiv:hep-ph/9910286].
  \bibitem{Barger:2001hy}
  V.~D.~Barger, D.~Marfatia and B.~P.~Wood,
  Phys.\ Lett.\ B {\bf 498}, 53 (2001)
  [arXiv:hep-ph/0011251].
   
\bibitem{BOREXINO}            
  G.~Ranucci {\it et al.}  [BOREXINO Collaboration],
  Nucl.\ Phys.\ Proc.\ Suppl.\  {\bf 91}, 58 (2001).

\bibitem{KamLAND} "Proposal for US Participation in KamLAND", March 1999,
http://kamland.lbl.gov/KamLAND.US.Proposal.pdf .             

\bibitem{thin} 
C. Lunardini, A.Yu. Smirnov, Nucl. Phys. {\bf B583} 260 (2000); 
E. K. Akhmedov,  Phys. Lett. {\bf B503} 133 (2001);  
O.  Yasuda,  Phys.Lett. {\bf B516} 111 (2001).   

\bibitem{Krastev:2001tv}
V. Barger, D. Marfatia,  hep-ph/0212126;  
G.L. Fogli, E. Lisi, A. Marrone, D. Montanino, 
A.  Palazzo, A.M. Rotunno, hep-ph/0212127; 
M. Maltoni, T. Schwetz, J.W.F. Valle, 
hep-ph/0212129;   
Abhijit Bandyopadhyay, Sandhya Choubey, Raj Gandhi, 
Srubabati Goswami, D.P. Roy, hep-ph/0212146;  
John N. Bahcall, M.C. Gonzalez-Garcia, Carlos Pena-Garay, hep-ph/0212147; 
H. Nunokawa, W.J.C. Teves, R. Zukanovich Funchal,  
hep-ph/0212202; P. de Holanda, A.Yu. Smirnov, hep-ph/0212270.    


  \bibitem{Dighe:1999id}
  A.~S.~Dighe, Q.~Y.~Liu and A.~Y.~Smirnov,
  arXiv:hep-ph/9903329.

  \bibitem{Dziewonski}
  A.~M.~Dziewonski and D.~L.~Anderson,
  Phys. Earth Planet.Interiors   {\bf 25}, 297 (1981).
   
\bibitem{Geller:2001ix}
  R.~J.~Geller and T.~Hara,
  arXiv:hep-ph/0111342.

\bibitem{bahcall}
A. V. Gruzinov, J. N. Bahcall,  Astrophys. J. {\bf 490} 437 (1997).  


\bibitem{BE}
  Encyclopaedia Britannica, 15th ed., Chicago, University, 1982.


\bibitem{uno}
M. Aglietta, W. Fulgione, O. Saavedra, G. Trinchero, 
Nucl. Instrum. Meth. {\bf A277} 17, (1989);  
Chang Kee Jun,  hep-ex/000504.  

\bibitem{param}
V.~K.~Ermilova, V.~A.~Tsarev and V.~A.~Chechin,
Kr. Soob. Fiz. {\it  Short Notices of the Lebedev Institute}, {\bf 5} 26 (1986),   
E. Kh. Akhmedov,  Yad. Fiz. {\bf 47} (1988) 475,    
P.I. Krastev, A.Yu. Smirnov, Phys. Lett. {\bf B226} :341-346, 1989,  
P. I. Krastev, A. Yu. Smirnov,   Mod. Phys. Lett. {\bf A6}:1001, 1991.  


\bibitem{stodolsky} A. Drukier and L. Stodolsky, Phys. Rev. {\bf D 30}, 
2295 (1984) L. Stodolsky, 
Phys. Rev. Lett. {\bf 34}, 110 (1975) [Erratum-ibid {\bf 34} 508 (1975)]. 

  \end{thebibliography}
  \end{document}